%% file: ijcai21.tex
\documentclass{article}
\usepackage{ijcai21}

\usepackage{times}
\usepackage{soul}
\usepackage{url}
\usepackage[hidelinks]{hyperref}
\usepackage[utf8]{inputenc}
\usepackage[small]{caption}
\usepackage{graphicx}
\usepackage{amsmath}
\usepackage{amsfonts}
\usepackage{amsthm}
\usepackage{booktabs}
\usepackage{algorithm}
\usepackage{algorithmic}
\usepackage{tikz}
\usepackage{pgfplots}
\pgfplotsset{width=7cm,compat=1.3}

\newtheorem{example}{Example}

\newtheorem{dfn}{Definition}
\newtheorem{lem}{Lemma}

\newtheorem{thm}{Theorem}

\title{Modeling Precomputation In Games Played Under Computational Constraints\footnote{To appear in IJCAI 2021.}}

\author{
Thomas Orton
\affiliations
University of Oxford
\emails
thomas.orton@cs.ox.ac.uk
}

\begin{document}

\maketitle

\begin{abstract}
  Understanding the properties of games played under computational constraints remains challenging. For example, how do we expect rational (but computationally bounded) players to play games with a prohibitively large number of states, such as chess? This paper presents a novel model for the precomputation (preparing moves in advance) aspect of computationally constrained games. A fundamental trade-off is shown between randomness of play, and susceptibility to precomputation, suggesting that randomization is necessary in games with computational constraints. We present efficient algorithms for computing how susceptible a strategy is to precomputation, and computing an $\epsilon$-Nash equilibrium of our model. Numerical experiments measuring the trade-off between randomness and precomputation are provided for Stockfish (a well-known chess playing algorithm). 
\end{abstract}

\section{Introduction}

While there is a firm theoretical understanding of how perfectly rational agents should play games in equilibrium, understanding the setting where agents have bounded rationality appears to be more challenging. For example, in theory, an agent playing a mini-max game such as chess always has an optimal move it can make. In practice, it may take a very long time to compute an optimal move, and it can be unclear how a rational agent should (or would) behave when it only has a finite amount of time to decide on a move. 

Early work on bounded rationality in game theory was motivated by explaining the behavior of agents playing repeated games of the prisoner's dilemma \cite{neyman-prisoners}, \cite{rubinstein-prisoners}, where agents with bounded rationality were modeled by finite automata. Related work \cite{Megiddo-bounded-rationality} extended this to the case where agents were modeled as Turing machines with a restricted number of internal states. As more general models of computation were explored, relations were drawn to well established complexity classes in theory of computation \cite{Papadimitriou-bounded-rationality}. These included explicit external utility costs for complex strategies \cite{ben-sasson-bounded-rationality},\cite{Halpern-bounded-rationality},\cite{Halpern-bounded-rationality-2}, or implicit discounting of utility for strategies which take longer to compute \cite{Fortnow-Santhanam-bounded-rationality}. Much of this line of work was concerned with whether Nash equilibria still exist in this setting, whether they are computable, and what they look like in limiting cases. For mini-max games in particular, a line of work tries to explain why increasing the mini-max search tree depth (i.e. spending more compute time per move) tends to improve agent performance \cite{chess-depth-vs-elo}. Under certain models increasing the search depth actually worsens the agent's quality \cite{beal-trees}, \cite{nau-trees}, \cite{pearl-trees}, while subsequent work has proposed alternative models to avoid this pathology. (see e.g. \cite{real-valued-minimax}). A drawback of much of the prior work mentioned, which is an indication of the challenge of modelling bounded rationality, is that either the models considered are often inflexible and only apply to toy problems (e.g. finite automata, idealized models for search trees, limiting behavior), or the results themselves are infeasible to compute (e.g. equilibrium is computable, but takes exponential time). In contrast, this paper aims to give a relatively flexible model, where some of its properties can be efficiently computed or approximately analyzed. 

Consider an extensive form game played between two players, where each player has a finite amount of total time during the game to make their moves. One simple yet under-explored property of these games in the time-constrained setting is the following: before the game, both players can typically spend a large amount of time practicing and preparing, and this time is typically much larger than the amount of time spent playing the game. During the game, if players encounter situations they have prepared for in advance, they are able to play very strong moves quickly from their preparation memory. Otherwise, they have to use their time-limited budget to compute a new move. This simple structure leads to interesting trade-offs. For example, if player 1 plays very deterministically, it will be easy for player 2 to prepare (precompute) strong moves against player 1 using the much larger compute resources available before the match. On the other hand, if player 1 plays more randomly to try and make the future of the game harder to predict, player 1 necessarily needs to play ``optimal" moves less frequently. Modeling just this precomputation trade-off leads to surprisingly rich behavior in time constrained games. For example, in chess (which will be used as a running example throughout the paper), human players often (a) prepare against other players by studying certain opening lines they expect to occur in a game (b) play the first few moves of the game quickly from an ``opening book" of memorized moves, and (c) intentionally randomize their strategies to make it harder for their opponent to prepare against them. In contrast, many modern chess engines do not explicitly exhibit some of the behavior described above; for example, they will deterministically play the ``best move" found within an allotted time, mirroring the idea that in the computationally unbounded setting, there is always an optimal move to play for a mini-max game like chess. In contrast, this paper will show that in the computationally bounded setting where players can precompute, it is perhaps more useful to think in terms of the ``best distribution of moves" to play. 

\textbf{Contributions:} We first give a novel and flexible formalism for modeling the precomputation aspect of two player, zero sum, perfect information extensive form games played under time constraints. Next, we give theoretical results establishing the importance of randomness of play in the computationally constrained setting. We show that one can efficiently compute how exploitable a fixed strategy is to precomputation, and the equilibrium precomputation strategy between two players. We then empirically demonstrate how useful precomputation can be in practical contexts, by exploring how susceptible Stockfish is to precomputation as a function of randomization. 

\section{Definitions}

\subsection{Background}

We consider two-player, zero sum, perfect information extensive form games, where the players alternate in turns. In order to provide maximum flexibility of the model, we use standard terminology for game playing which makes no explicit reference to computational constraints. However, we implicitly think of player policies $\sigma_i$ as being implemented by some computationally constrained algorithm (e.g. alpha-beta search), and the game being played under some time limit. We will give some concrete examples shortly.

A two-player game consists of a set of \textbf{histories} $H$, a subset of \textbf{terminal histories} $Z \subset H$,  a \textbf{utility function} $u:Z \rightarrow [0,1]$ defined on terminal histories $h \in Z$, finite sets of possible actions $A(h)$ for each history $h \in H \setminus Z$, and a \textbf{transition function} $\pi$ which takes as input a history $h \in H \setminus Z$, an action $a \in A(h)$, and returns a unique history $h'=\pi(h,a)$. 

A \textbf{behavioral strategy} (policy) for player $i \in \{1,2\}$ consists of a function $\sigma_i$ defined on $H$, where $\sigma_{i}(h) \in \Delta(A(h))$\footnote{Here $\Delta$ is the probability simplex.}, i.e. $\sigma_i$ maps histories to distributions over actions, where $\sigma_{i}(h)(a)$ is the probability that player $i$ plays action $a \in A(h)$ when at history $h$. A \textbf{player function} $P:H \rightarrow \{1,2,c\}$ indicates which player's turn it is to choose an action at history $h$. If $P(h)=c$, then $h$ corresponds to a placeholder chance node (i.e. chance player) and an action $a \in A(h)$ is chosen according to a fixed probability distribution $\sigma_{c}(h)$. 

 For notational reasons, we associate each history $h \in H$ with the unique ordered sequence $(a_1,\dots,a_{k})$ of actions which must be played to reach $h$ beginning from the starting history $\emptyset \in H$, and we say $h$ has length $|h|=k$. For $i \in \{1,\dots,k\}$ we denote $h_i=a_i$, and denote by $h_{:i}$ the history reached by playing $a_1,\dots, a_{i}$, so in particular we have $h_{:i}=\pi(h_{:i-1},a_{i})$. We say $h \in H$ is a prefix of $h'$ if $\exists i$ s.t. $h=h'_{:i}$, and we write this as $h \leq h'$. The \textbf{successor function} $Succ(h):H \rightarrow 2^{H}$ (here $2^{H}$ is the power set of $H$) maps a history $h$ to the set of histories $H' \subset H$ where player $P(h)$ next gets to have a turn.
 Unless explicitly stated, we will consider games where \textbf{player 1 and 2 alternate in turns} with no chance nodes, i.e. $P(h)=1$ if $|h|$ is even, $P(h)=2$ otherwise. 

 Given a \textbf{policy profile} $\sigma = (\sigma_1,\sigma_2)$ for both players, the probability of history $h'$ occurring when starting from history $h \leq h'$ is therefore  $\pi^{\sigma}(h,h')=\prod_{i=|h|}^{|h'|-1} \sigma_{P(h'_{:i})}(h'_{:i})(h'_{i+1})$, and $0$ when $h \not \leq h'$. We define the probability of history $h$ as $\pi^{\sigma}(h)=\pi^{\sigma}(\emptyset,h)$. We say that the expected value of the game for player $1$ when starting from history $h \in H$ is $u_{1}^{\sigma}(h):=\sum_{h' \in Z} \pi^{\sigma}(h,h') u(h')$, and $u_{2}^{\sigma}(h):=1-u_{1}^{\sigma}(h)$ for player $2$. The expected value of the game for player $1$ is $u^{\sigma}_{1}:=u^{\sigma}_{1}(\emptyset)$.

\subsection{Algorithmic Strategies and Precomputation}

Precomputation strategies are intended to capture the idea that, before a game, we can study some subset $S \subset H$ of game histories and plan in advance what to play for these histories. For example, in the case of chess, we could look at some board positions $S \subset H$, and see which moves a powerful chess engine $\sigma_{pre}$ would recommend in these positions after running for an hour. In the actual game, if we ever end up in a position $h \in S$, we can immediately play $\sigma_{pre}(h)$ from a lookup table (without spending any time computing moves). Otherwise, if $h \not \in S$, we can just play our usual policy $\sigma_{i}(h)$ (where we would need to spend compute time during the game).

\begin{dfn}\label{def:precomp-strategy}
Given policies $\sigma_{pre},\sigma_{i}$, a $(\sigma_{pre},\sigma_{i},B)$ \textbf{precomputation strategy} with memory budget $B$ is a policy $\tilde{\sigma}_{i}$ such that:

\begin{enumerate}
    \item There exists a memorization set $S \subset H$, where $|S| \leq B$ and $S$ is prefix closed (i.e. if $h \in S$ and $h'<h,P(h')=P(h) \implies h' \in S$). 
    
    \item $h \in S \implies \tilde{\sigma}_{i}(h)=\sigma_{pre}(h)$.
    
    \item $h \not \in S \implies \tilde{\sigma}_{i}(h)=\sigma_{i}(h)$.
    
\end{enumerate}
\end{dfn}

For convenience, we will always think of modeling policies $\sigma_{pre}$ as taking $0$ time per move. One justification for considering prefix closed memorization sets is that if a player chooses to memorize history $h \in H$, it is natural for them to have also considered the histories $h'<h$ leading up to $h$. We now consider the following meta-game: before a match between player $1$ and player $2$, which takes place in a time-constrained setting, each player $i$ can spend time preparing for the match by specially choosing a prefix-closed subset of histories $S \subset H$, and constructing a precomputation strategy $\tilde{\sigma}_i$ which plays strong moves according to some policy $\sigma_{pre}$ when in this set. We assume that there is some limiting factor on how many moves each player can memorize for a particular game. In practice, this may be enforced by e.g. a maximum memory budget $B$, or some limited capacity to memorize moves. We model this by penalizing each player by a linear factor depending on the size of the memorization set of their precomputation strategy. The theory in Section 3 is relatively robust to other choices of penalty functions, but we found the linear penalty led to the most efficient algorithms in Section 4. \footnote{However, one can still find polynomial time algorithms for other natural penalty functions such as a hard limit on the memorization set size.} For more concrete intuition of the formalism one can consider chess: a player can specially prepare a list of opening moves before a tournament, but their capacity to memorize chess lines for a particular match is limited, and their choice invariably depends on the opening move choices prepared for by their opponent as well. This results in an evolving ``meta game" of the best opening lines to play.

\begin{dfn}\label{def:meta-precomp-game}
(The meta-precomputation game) Let $Pre(\sigma_{pre},\sigma_i)$ denote the set of all $(\sigma_{pre},\sigma_i,K)$ precomputation strategies for all $K \in \mathbb{N}$, and denote $\Delta(Pre(\sigma_{pre},\sigma_i))$ by the set of all mixed strategies on $Pre(\sigma_{pre},\sigma_i)$. For $\tilde{\sigma} \in Pre(\sigma_{pre},\sigma_i)$, let $|\tilde{\sigma}|$ denote the size of the memorization set used by precomputation strategy $\tilde{\sigma}$. Consider a meta-game where each player $i$ instantaneously chooses a mixed precomputation strategy $\Delta\tilde{\sigma}_i \in \Delta(Pre(\sigma_{pre},\sigma_i))$, and the outcome utility of the game is

\begin{align*}
\tilde{u}^{(\Delta\tilde{\sigma}_1,\Delta\tilde{\sigma}_2)}:=\tilde{u}_{1}^{(\Delta\tilde{\sigma}_1,\Delta\tilde{\sigma}_2)}\\
=\mathbb{E}_{\tilde{\sigma}_1\sim \Delta\tilde{\sigma}_1, \tilde{\sigma}_2 \sim \Delta\tilde{\sigma}_2}\left[u_{1}^{(\tilde{\sigma}_1,\tilde{\sigma}_2)}-\lambda_1 |\tilde{\sigma}_1|+\lambda_2 |\tilde{\sigma}_2|\right]\\
\end{align*}

And likewise $\tilde{u}_2=1-\tilde{u}_{1}$. For $\lambda_1,\lambda_2 \in \mathbb{R}^{+}$, $\Delta \tilde{\sigma}_i\in \Delta(Pre(\sigma_i,\sigma_{pre}))$ for $i \in \{1,2\}$ is an $\epsilon$-Nash equilibrium if 

\begin{align*}
    \tilde{u}^{(\Delta\tilde{\sigma}_1,\Delta\tilde{\sigma}_2)}+\epsilon &\geq \sup_{\Delta\tilde{\sigma}_1' \in \Delta(Pre(\sigma_{pre},\sigma_1))}  \tilde{u}^{(\Delta\tilde{\sigma}_1',\Delta\tilde{\sigma}_2)}\\ 
    \tilde{u}^{(\Delta\tilde{\sigma}_1,\Delta\tilde{\sigma}_2)}-\epsilon&\leq  \inf_{\Delta\tilde{\sigma}_2' \in \Delta(Pre(\sigma_{pre},\sigma_2))}  \tilde{u}^{(\Delta\tilde{\sigma}_1,\Delta\tilde{\sigma}_2')}\\
\end{align*}

When this is an $\epsilon=0$ Nash equilibrium, we omit the $\epsilon$ and say that $v_{Nash}=\tilde{u}^{(\Delta\tilde{\sigma}_1,\Delta\tilde{\sigma}_2)}$ is a Nash equilibrium value of the game.

\end{dfn}

\subsection{Examples}

We now give two explicit examples of how we might model games with computational constraints in this framework:

\begin{example}
(Chess with a time limit per move): Suppose we are playing chess, so $h \in H$ corresponds to the prior moves made by each player, $A(h)$ contains all legal moves for player $P(h)$ in state $h$, and $u_{1}(z)=1$ if black is checkmated for $z \in Z$, $0$ if white is checkmated by black, and $\frac{1}{2}$ for a draw. We think of $\sigma_{1},\sigma_{2}$ as chess playing algorithms which each have $1$ second per move (concretely, $\sigma_{1},\sigma_{2}=Stockfish(1)$, Stockfish\footnote{Stockfish is a popular open source chess playing algorithm.} with a time limit of $1$ second per move). If $\tilde{\sigma}_{1}$ is a $(\sigma_{pre},\sigma_{1},B)$ precomputation strategy for white with memorization set $S$, and $\sigma_{pre}$ is a chess playing algorithm which spends 1 hour precomputing per move (concretely, $Stockfish(3600)$), then when $h \in S$, $\tilde{\sigma}_{1}(h)$ plays $Stockfish(3600)(h)$ instantaneously; otherwise $\tilde{\sigma}_{1}(h)$ runs $Stockfish(1)$ on board state $h$. 
\end{example}

\begin{example}\label{exmple:chess-total-time-limit}
(Chess with a total time limit per game): Consider the previous example, except that instead of having a time limit per move, we now have a time limit of $1$ hour for the entire game. Move histories $h \in H$ are now augmented to include the time each player took to make their corresponding moves, and the terminal set $Z$ now includes states where players have run out of time (in which case the opposing player wins).
\end{example}

Note that allowing the time an algorithm $\sigma_{1}$ takes to depend on the entire history of moves $h$ (instead of just the chess board state) allows us to capture the behavior of algorithms whose computation depends on what they computed in the past; for example, chess playing algorithms often keep a lookup table of moves and positions they have considered during prior moves, in order to speed up future computations. 

\section{When Do Good Precomputation Strategies Exist?}\label{section:existence}

In this section, we will study how susceptible a fixed policy $\sigma_i$ is to precomputation as a function of the randomness of the policy. For simplicity, we will make the definitions from the perspective of player 1 (but the analogous statements hold for player 2). Without loss of generality, we assume that $\forall h \in Z, P(h)=1$, i.e. the game always ends on player 1's turn. We can do this by adding a sentinel history to every terminal state where this is not the case.  

To relate precomputation strategies to randomness of play, we show how to explicitly construct a precomputation strategy whose size depends on how randomly $\sigma_2$ plays with respect to $\sigma_{pre}$. This construction then serves as a lower bound on the quality of the best precomputation strategy. The first key observation we make is the following: suppose the game reaches a state $h \in H$ such that player 1 now has significant advantage against player 2. Then it should be the case that player 1 can play their normal strategy $\sigma_1$ against $\sigma_2$, without any precomputation, and convert this game to a win. Therefore, one way of constructing a precomputation strategy is the following: have player 1 play precomputed moves from $\sigma_{pre}$ against $\sigma_2$ until a significant advantage (of value $v$) is gained, and then continue to play normal moves from $\sigma_1$ afterwards. If $\sigma_{pre}$ is a much stronger policy than $\sigma_2$, then we expect to not need to play too many moves until an advantage is reached. We can make this observation more concrete by considering the domain of chess: player 1 can play very strong memorized chess engine moves until player 1 has a substantial advantage (e.g. has more pieces on the board than player 2). Player 1 can then play normally to convert this advantage to a win. We formalize this idea with the following definition:

\begin{dfn}
Given policy profile $\sigma=(\sigma_1,\sigma_2)$, a policy $\sigma_{pre}$ and $v \in [0,1]$, the set histories where player 1 first gets an advantage $\geq v$ is

\begin{align*}
    S^{\sigma}(v) :=
    \{h' \in H |P(h')=1 \land u_{1}^{\sigma}(h') \geq v \\ \land \not \exists  h < h'\text{ s.t. }u_{1}^{\sigma}(h)\geq v \land P(h)=1\}\\
\end{align*}

We define a distribution $\mathcal{S}^{\sigma,\sigma_{pre}}(v)$ on $S^{\sigma}(v)$ via

$$\Pr_{\mathcal{S}^{\sigma,\sigma_{pre}}(v)}[h] := \frac{\pi^{(\sigma_{pre},\sigma_2)}(h)}{P_{norm}(v)}$$

where the normalizing constant

$$P_{norm}(v)=\sum_{h \in S^{\sigma}(v)} \pi^{(\sigma_{pre},\sigma_2)}(h)$$

is the probability of reaching any $h \in S^{\sigma}(v)$ when $\sigma_{pre}$ plays against $\sigma_2$. 

\end{dfn}

\begin{example}
For intuition, if $\sigma_{pre}$ is a much stronger policy than $\sigma_2$, we expect $P_{norm}(v)$ to be close to $1$. In the running example of chess, if $\sigma_{pre}=Stockfish(1000)$ and $\sigma_1,\sigma_2=Stockfish(1)$, we have $P_{norm}(v)\geq P[\text{Stockfish(1000) as white checkmates Stockfish(1) as black}]$ for all $v \leq 1$, because $u^{\sigma_1,\sigma_2}_{1}(h)=1$ in every terminal history $h$ where white has checkmated black.
\end{example}

We now observe that if the entropy $H(\mathcal{S}^{\sigma,\sigma_{pre}}(v))$ of the distribution $\mathcal{S}^{\sigma,\sigma_{pre}}(v)$ is small, then we do not need to memorize too many states to construct the precomputation strategy which memorizes moves until an advantage $v$ is reached. Since we get a value of at least $v$ whenever we end up in $h \in S^{\sigma}(v)$, we expect a utility of approximately $v P_{norm}(v)$. We can quantify this observation with the following Theorem:

\begin{thm}\label{thm:rand-vs-precomp}
Given $\sigma=(\sigma_1,\sigma_2)$ and a policy $\sigma_{pre}$, suppose that the maximum number of moves per game is $L:=\max_{h \in H} |h|$. 
Then $\forall v \in [0,1],\epsilon \in (0,1)$, there exists a $(\sigma_{pre},\sigma_1,L(1-\epsilon)e^{H(\mathcal{S}^{\sigma,\sigma_{pre}}(v))/\epsilon})$ precomputation strategy $\tilde{\sigma}_1$ for player 1 such that

$$u^{(\tilde{\sigma}_1,\sigma_2)}_{1} \geq (1-\epsilon) v P_{norm}(v)$$

where $H$ is the (base $e$) entropy of the distribution $\mathcal{S}^{\sigma,\sigma_{pre}}(v)$.

\end{thm}

For any fixed $v \in [0,1],\epsilon \in (0,1)$, Theorem \ref{thm:rand-vs-precomp} therefore gives us a lower bound on how exploitable a policy $\sigma_2$ is to precomputation. As expected, there is a direct trade-off between the size of the precomputation set required to exploit $\sigma_2$, and how randomly $\sigma_2$ plays against some fixed policy $\sigma_{pre}$.

We now make the following observation: suppose that $\Delta \tilde{\sigma}_{2}^{*}$ is a Nash equilibrium strategy for player 2 in the meta-precomputation game. If this equilibrium has a favourable equilibrium value for player 2, then it follows from Theorem \ref{thm:rand-vs-precomp} that $\mathcal{S}^{\sigma,\sigma_{pre}}(v')$ must have high entropy. For the sake of contradiction, suppose that this were not the case: it would then be possible to construct a small precomputation strategy for player 1 which performs well against $\Delta \tilde{\sigma}_{2}^{*}$ but has low precomputation penalty, contradicting the assumption that $\Delta \tilde{\sigma}_{2}^{*}$ has a favourable equilibrium value for player 2. By following this proof by contradiction logic precisely, we obtain the following Theorem. We simplify the statement by considering the case where $\sigma_{pre}$ always wins against $\Delta \tilde{\sigma}_{2}^{*}$ in the original game.

\begin{thm}\label{thm:precomp-game-entropy-bound}
Suppose $\Delta \tilde{\sigma}_2^{*}$ is a Nash equilibrium strategy for player $2$ with value $v$ in the meta-precomputation game. Suppose that $u_{1}^{(\sigma_{pre},\Delta \tilde{\sigma}_2^{*})}=1$,\footnote{i.e. $\sigma_{pre}$ always beats $\Delta \tilde{\sigma}_2^{*}$ in the original game, where we implicitly associate the mixed strategy $\Delta \tilde{\sigma}_2^{*}$ with a corresponding behavior strategy in the original game by Kuhn's theorem.} the maximum number of moves in the game is $L$, and that the precomputation penalty term for player 1 is $\lambda_1>0$. Then $\forall v' \in (v+0.01,1)$, we have that

$$H\left(\mathcal{S}^{(\sigma_1,\Delta \tilde{\sigma}_2^{*}),\sigma_{pre}}(v')\right) \geq (v'-v-0.01)\left(\log\left(\frac{1}{\lambda_1 L}\right)-5\right)$$
\end{thm}

For any fixed $v' \in (v+0.01,1)$, Theorem \ref{thm:precomp-game-entropy-bound} therefore quantifies a lower bound on how randomly equilibrium policies play in the precomputation model. The bound tells us that the better the strategy $\Delta \tilde{\sigma}_2^{*}$ is for player 2 (the smaller the equilibrium value $v$ is), the more randomly player 2 must play. In contrast to e.g. mini-max games in the computationally unbounded setting where optimal strategies are deterministic, optimal strategies in this model are inherently randomized. 

\section{Efficiently Finding Precomputation Strategies And Their Equilibria}\label{section:finding_precomp}

Suppose we are given oracle access to algorithmic strategies $\sigma_1,\sigma_2,\sigma_{pre}$. Concretely, we could have $\sigma_1,\sigma_2,\sigma_{pre}=Stockfish$ with different time per move settings. We might believe there is a good precomputation strategy for $\sigma_1$ against $\sigma_2$, but it is perhaps unclear how to find such a strategy. Suppose $\tilde{\sigma}_{1}$ is an optimal precomputation strategy for player 1 against $\sigma_2$, with memorization set $S$. The first observation is that we can assume $h \in S \implies \pi^{(\sigma_{pre},\sigma_2)}(h) \geq \lambda_1$; otherwise, player $1$ could strictly increase $\tilde{u}^{(\tilde{\sigma}_1,\sigma_2)}_{1}$ by removing $h$ and any descendants of $h$ from $S$. This is because the cost of including $h$ and any descendants of $h$ in $S$ is at least $\lambda_1$, but the gain in utility from any histories which pass through $h$ when precomputing up to $h$ is at most $\pi^{(\sigma_{pre},\sigma_2)}(h)$. The second observation is that we can conclude $|S| \leq \frac{L+1}{\lambda_1}$ from the following Lemma:

\begin{lem}\label{Lemma:bound-on-set-size}
$|\{h \in H: \pi^{\sigma_1,\sigma_2}(h) \geq p\}|\leq \frac{L+1}{p}$, where $\sigma_1,\sigma_2$ are any policies and the maximum history length of the game is $L:=\max_{h \in H} |h| $. 
\end{lem}

Thus finding the optimal precomputation strategy for player 1 against $\sigma_2$ is equivalent to finding the optimal subtree (memorization set) contained within a bounding tree of size $\frac{L+1}{\lambda_1}$ to memorize. By using standard Chernoff bounds to sample the value of the game at the leaves of this bounding tree, we can compute the optimal subtree with standard dynamic programming techniques. The following Theorem makes these ideas precise, and the detailed algorithm (Algorithm 1) can be found in the technical appendix.

\begin{thm}\label{thm:find-precop-fixed}
For any $\epsilon,\delta >0$, and precomputation penalty factor $\lambda_1>0$ for player 1, suppose we are given a policy profile $\sigma=(\sigma_1,\sigma_2)$ and a policy $\sigma_{pre}$ as constant-time oracles. Suppose further that $\exists \tilde{\sigma}_1 \in Pre(\sigma_{pre},\sigma_1)$ such that $\tilde{u}_{1}^{(\tilde{\sigma}_1,\sigma_2)}\geq v$. Then with probability at least $1-\delta$, Algorithm 1 will return a $\tilde{\sigma}_1' \in Pre(\sigma_{pre},\sigma_1)$ such that $\tilde{u}_{1}^{(\tilde{\sigma}_1',\sigma_2)}\geq v-\epsilon$. Moreover, Algorithm 1 runs in time $\mathcal{O}\left(A^2 \frac{L^2}{\lambda_1\epsilon^2}  \ln(\frac{A^2L}{\lambda_1})\right)$, where $L=\max_{h \in H} |h|$ and $A=\max_{h \in H} |A(h)|$. 
\end{thm}

\subsection{Computing Equilibria}

In the previous setting, we imagined the opposing player as fixed, while the precomputing player optimizes the best set of game states to memorize in order to exploit the opposing player. In reality, the opposing player reacts to this precomputation by preparing their own lines of play in anticipation to counter this precomputation. In this subsection we will show how an $\epsilon-$Nash equilibrium for the meta-precomputation game can be efficiently found, by reducing the problem to a form which can be solved by the well-studied method of Counter Factual Regret Minimization (CFR) \cite{regretminCRF1}. To compute an equilibrium of the meta precomputation game, we first observe (similar to the previous subsection) that we can ignore low probability histories $h \in H$ where the possible gain in utility of precomputing at $h$, regardless of the opposing player's strategy, is always upper bounded by the precomputation penalty $\lambda_i$. For example, player 2 only has an incentive to precompute on the set $W_{2}=\left\{h \in H|  \sup_{\Delta \tilde{\sigma}_{1} \in \Delta Pre(\sigma_{pre},\sigma_{1})} \pi^{\Delta \tilde{\sigma}_{1},\sigma_{pre}}(h) \geq \lambda_2\right\}$. If we denote the set of high probability histories by $W$, then we can bound the size of $W$ by $|W| \leq \mathcal{O}\left(A^2(L+2)^2(\frac{1}{\lambda_1}+\frac{1}{\lambda_2})\right)$ (Lemma 3 in technical appendix).  

The second observation is that the meta-precomputation game $G$ can be viewed as an equivalent extensive form game $G'$ with imperfect information and perfect recall. At each game state $h \in H$, we can imagine that instead of players choosing a distribution of actions in $A(h)$ to play, they choose whether to play an action from distribution $\sigma_{pre}(h)$ (i.e. continue precomputing) or distribution $\sigma_{i}(h)$ (stop precomputing); however, if a player chooses to play from distribution $\sigma_{pre}(h)$, they must pay a cost of $\lambda_i$ corresponding to the precomputation penalty. We must re-weight this penalty so that it is not discounted by the probability of reaching board state $h$. Players cannot observe whether their opponent chose to play from distribution $\sigma_{pre}(h)$ or $\sigma_{i}(h)$, but they do see the action which was sampled from the distribution their opponent chose. The memorization set for a precomputation strategy then naturally corresponds to the set of histories where a player chooses to play from distribution $\sigma_{pre}(h)$, and so there is a bijective correspondence between \textit{pure strategies} in game $G$, and \textit{pure strategies} in game $G'$. Using the equivalence between behavior strategies and mixed strategies for extensive form games with perfect recall (Kuhn's Theorem), we then get an equivalence between behavior strategies in game $G'$, and mixed strategies in game $G'$. The precise definition of $G'$, and a proof of its equivalence, can be found in the supporting technical appendix (Lemma 4). Thus if we can find an $\epsilon$-Nash equilibrium of behavior strategies in game $G'$, we can find an $\epsilon$-Nash equilibrium of mixed strategies for the meta-precomputation game. Moreover, finding an $\epsilon$-Nash equilibrium for behavior strategies for this new game can be solved efficiently by CFR if we apply previous insights. In particular, we remove low probability histories $H\setminus W$ from the game, and estimate the value of terminal histories with random roll-outs. The following theorem makes these details precise, and the algorithm (Algorithm 2) can be found in the technical appendix.

\begin{thm}\label{thm:CFR-on-game}
For any $\epsilon>0,\delta>0$ and precomputation penalty factors $\lambda_1,\lambda_2>0$, suppose we are given a policy profile $\sigma=(\sigma_1,\sigma_2)$ and a policy $\sigma_{pre}$ as constant-time oracles. Then Algorithm 2 returns an $\epsilon$-Nash equilibrium to the meta-precomputation game with probability at least $1-\delta$. Moreover, the total number of CFR iterations is bounded by $T=\mathcal{O}\left(\frac{|W|^2}{\epsilon^2}\right)$, and the total runtime is bounded by 

$$\frac{1}{\epsilon^2}\left(A^2L^2\left(\frac{1}{\lambda_1}+\frac{1}{\lambda_2}\right)\right)^3$$

$$+\frac{1}{\epsilon^2}A^2L^3\left(\frac{1}{\lambda_1}+\frac{1}{\lambda_2}\right)\ln\left(AL\left(\frac{1}{\lambda_1}+\frac{1}{\lambda_2}\right)/\delta\right)$$

\end{thm}

While these bounds show that in theory one can approximate the Nash equilibrium in polynomial time, in practice it is possible that the equilibrium can be reached after fewer than  $T=\mathcal{O}\left(\frac{|W|^2}{\epsilon^2}\right)$ iterations of CFR. In particular, at iteration $T'<T$, one can use Theorem \ref{thm:find-precop-fixed} to verify whether the $\epsilon$-Nash equilibrium condition holds by computing the optimal precomputation strategy value against the current opponent strategy. One can use different variants of CFR which are designed to converge faster in practice depending on the specific structure of the game in question.

\section{Experiments}

While numerically evaluating CFR to compute equilibria in games is well-studied, finding the correct heuristic variant of CFR to make Algorithm 2 converge quickly for a particular game in practice is a research question on its own, and beyond the scope of this paper. However, the susceptibility of real-world strategies to precomputation, and how useful precomputation can be as a function of the randomness of an opposing strategy, is both relatively unexplored and also feasible to compute in practice without further optimization. We therefore choose to focus on this aspect, and propose an experiment to numerically explore the popular chess engine Stockfish's susceptibility to precomputation using Algorithm 1. Specifically, we choose $\sigma_{pre}$ to be Stockfish with $50$ms per move, while $\sigma_1$ and $\sigma_2$ play as Stockfish with $10$ms per move. For example, Algorithm 1 will transform white=Stockfish(10ms) into a new algorithm white=Stockfish'(10ms), which still takes at most 10ms per move, uses slightly more memory, and performs substantially better against black=Stockfish(10ms) if black does not sufficiently randomize. In particular, we explore how varying the randomness of the policy affects its susceptibility to precomputation. To do this, we introduce a randomness parameter $r$, and have a policy play more randomly as $r$ increases. As $r \rightarrow 0$, $\sigma_{i}$ will play only the best moves found by Stockfish in the required time period. As $r \rightarrow \infty$, $\sigma_{i}$ will play uniformally between available moves. If player $i$ is the fixed opponent, we set $\sigma_{i}(h)=\text{Softmax}(Stockfish(h)/r)$, where $\text{Softmax}$ is the softmax function and $Stockfish(h)$ returns a vector of the center pawn (cp) scores of available moves.\footnote{A center pawn score of 100 corresponds to having an advantage of one center pawn.} First we fix the default policies for white ($\sigma_1$,$\sigma_{pre}$ with $r=10^{-6}$), and vary $r$ for black ($\sigma_2$), computing the optimal precomputation value against black for each level of randomness. Then we repeat the experiment for black precomputing against white. We set $\lambda_{1}=\lambda_{2}=10^{-5}$ in these experiments, and plot the precomputed strategy utility (without precomputation penalty) and memorization set size for varying levels of randomness. In order to keep the computation requirements modest, only the top $K=2$ moves of the Stockfish engine were considered at each board position. Instead of sampling the game value, a conservative cp bound was used to approximate when either player had a decisive advantage. Further experimentation details can be found in the appendix, and the full code and technical appendix can be found on Github.\footnote{\href{https://github.com/Thomas-Orton/chess-precomputation}{https://github.com/Thomas-Orton/chess-precomputation}.}

\begin{figure}[t]
\centering
\includegraphics[width=1\columnwidth]{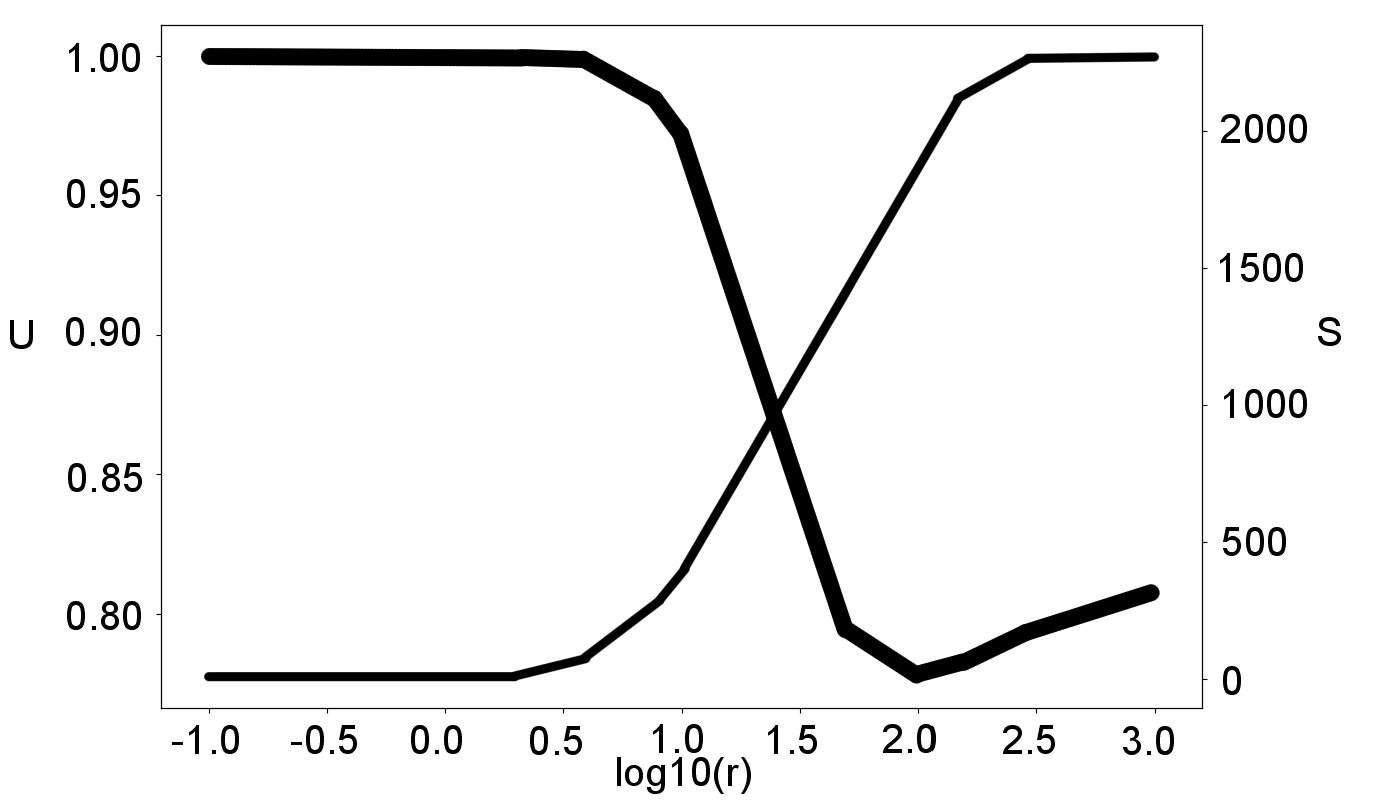} 
\includegraphics[width=1\columnwidth]{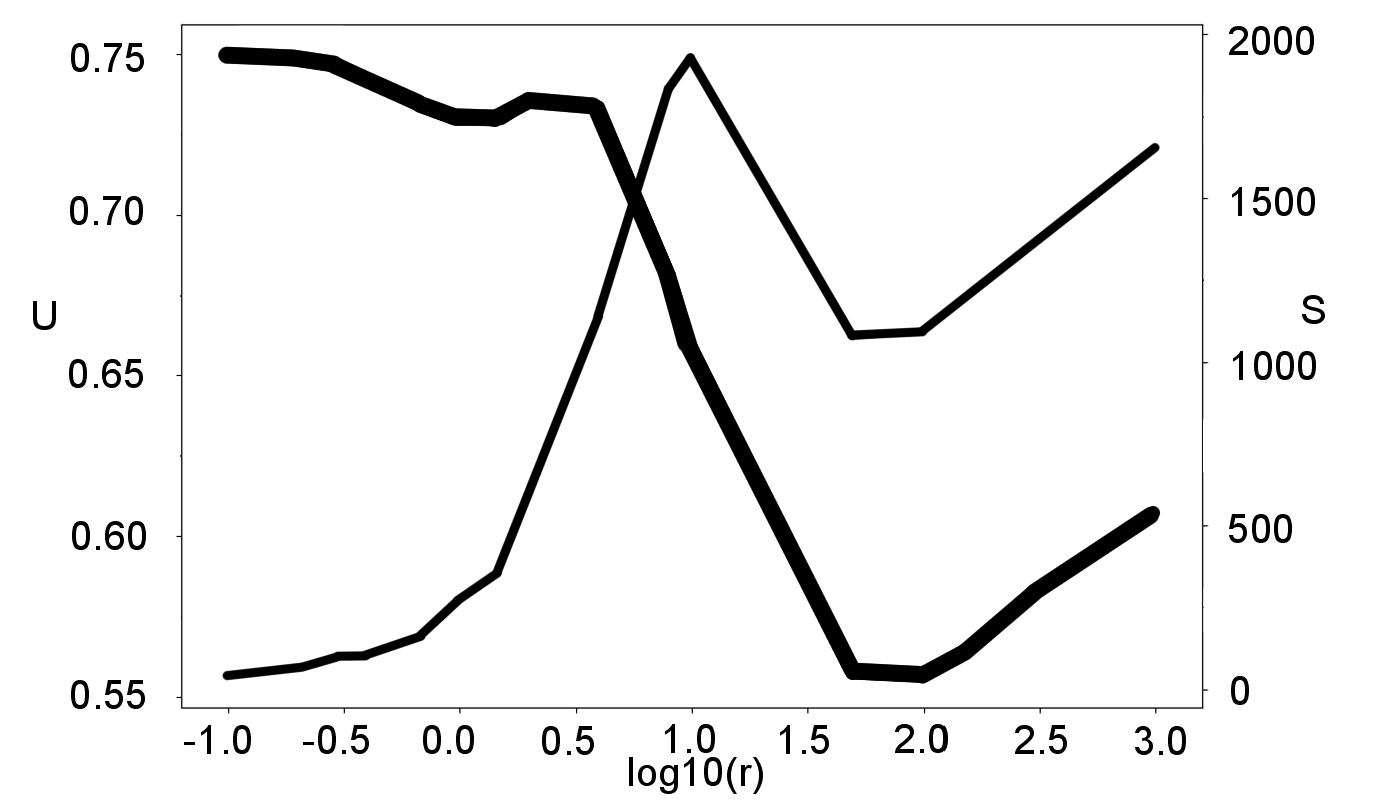}
\caption{Precomputation effectiveness as a function of player randomness; Top: white as precomputing player. Bottom: black as precomputing player. The utility $U$ of the precomputation strategy for the precomputing player, ignoring the precomputation memorization penalty (thicker line, left axis) and the size $S$ of the optimal precomputation set (thinner line, right axis) are plotted against $\log_{10}(r)$.}.
\label{fig1}
\end{figure}

We see from the results that when either white or black is precomputing, precomputation is particularly effective when the opposition plays predictably ($r$ is small). As the randomness parameter increases, precomputation effectiveness decreases, and more states need to be memorized (the precomputation set size increases). In both plots, there is a transition point at $r=100$ (corresponding to one center pawn) where the gain in playing more randomly against precomputation is offset by the loss in playing suboptimally. This suggests an interesting trade-off between playing too predictably and playing the perceived best moves available.

Note that white is able to achieve generally higher precomputation values against black. This can be explained by two observations. The first is that white is generally considered to have an advantage against black, and so Stockfish with $50$ms as black does not always win against Stockfish with $10$ms as white. The second is that black moves second, which means that black must remember more move variations than white to get to the same precomputation game depth in the game tree. This suggests that the first moving player has an inherent advantage when it comes to precomputation.

\section{Conclusion}

An unsolved problem in computer science is understanding how computationally bounded agents play games. We contribute to the understanding of this problem by presenting a novel formalism for modeling the precomputation aspect of computationally constrained agents, and show that in contrast to the computationally unconstrained setting, randomization plays an essential role for constructing effective strategies. This result is perhaps in opposition to the conventional heuristic of (deterministically) playing the strongest move found within the available time period. Moreover, we showed that optimal strategies in this model can be computed efficiently, and presented empirical results which suggest that precomputation strategies are practically useful in real-world time constrained games such as chess. Our model is flexible, and is essentially equivalent to charging a player depending on which strategy ($\sigma_i$ or $\sigma_{pre}$) they play from in each history. However, the model also has a fundamental structure in connection to the following question in computational complexity theory. Consider a game where there is a time limit of 1 second per move. Informally speaking, suppose we believe there is an algorithm $\sigma_1$ which achieves value at least $v$ against any other algorithm $\sigma_2$ which also takes 1 second per move (whether such a $\sigma_1$ exists is a highly non-trivial problem).\footnote{Here we are informally ignoring the details of program size.} Because a precomputation strategy $\tilde{\sigma}_{2} \in Pre(\sigma_{pre},\sigma_2)$ takes no more time than one second per move, it follows that $\sigma_1$ must also have value at least $v$ against $\tilde{\sigma}_2$, i.e. $\sigma_1$ must be robust against other algorithms hard-coding constants into their strategy and having non-uniform advice. Exploring this relationship in detail is an area for future work, but the observations about randomness being essential for competitive play are expected to carry over.

\bibliographystyle{named}
\bibliography{ijcai21.bib}

\include{appendix}

\end{document}

%% file: appendix.tex
\appendix

\section{Technical Appendix}
\textbf{Proof of Theorem} \ref{thm:rand-vs-precomp}.
\begin{proof}
First we need the following simple Lemma:

\begin{lem}\label{lem:low-entropy-implies-conc-mass}
Let $D$ be a discrete distribution with probabilities $p_1 \geq p_2 \geq \dots$. $\forall \gamma \in (0,1), z \in \mathbb{N}$, it holds that if $ z\geq \gamma e^{H(D)/(1-\gamma)}$, then $\sum_{i=1}^{z} p_i \geq \gamma$.
\end{lem}

Order the histories $h_1,\dots \in S^{\sigma}(v)$ descending by probability with respect to $\pi^{(\sigma_{pre},\sigma_2)}$, with probabilities $p_1 \geq p_2\geq \dots$ respectively. Then setting $\gamma=1-\epsilon$ in Lemma \ref{lem:low-entropy-implies-conc-mass} gives us 

$$\sum_{i=1}^{z} \pi^{(\sigma_{pre},\sigma_2)}(h_i) \geq (1-\epsilon) P_{norm}(v)$$

Where $z=\lceil(1-\epsilon)e^{H(\mathcal{S}^{\sigma,\sigma_{pre}}(v))/\epsilon}\rceil$. Now consider the precomputation strategy $\tilde{\sigma}_1$ which memorizes all histories in $S:=\{h \in H: \exists i \in [z] \text{ s.t. } h < h_i\}$, where $|S| \leq L z$. Then  $\forall i \in [z]$, we have $\pi^{(\sigma_{pre},\sigma_2)}(h_i)=\pi^{(\tilde{\sigma}_{1},\sigma_2)}(h_i)$. Since $h_i \not \leq h_j$ for $i \not = j$, and $\forall i \in [z], u^{(\sigma_1,\sigma_2)}(h_i) \geq v$, we have

$$u^{(\tilde{\sigma}_1,\sigma_2)} \geq \sum_{i = 1}^{z} \pi^{(\sigma_{pre},\sigma_2)}(h_i) u^{(\sigma_1,\sigma_2)}(h_i)\geq (1-\epsilon)v P_{norm}(v)$$

\end{proof}

\textbf{Proof of Lemma} \ref{lem:low-entropy-implies-conc-mass}.

\begin{proof}
Suppose $\sum_{i=1}^{z} p_i < \gamma$. Then we have

\begin{align*}
    H(D) &\geq \sum_{i=z+1}^{|D|} p_i \log\left(\frac{1}{p_i}\right)\\ 
    &\geq \sum_{i=z+1}^{|D|} p_i \log\left(\frac{1}{p_z}\right)\\ 
    & > (1-\gamma) \log\left(\frac{1}{p_z}\right)\\
    &\geq  (1-\gamma) \log\left(\frac{1}{\gamma/z}\right)\label{ineq:pz}\\ 
\end{align*}

Where the last inequality follows because $p_z \leq \gamma/z$. Solving gives

\begin{align*}
z & < \gamma e^{H(D)/(1-\gamma)}
\end{align*}

and the result follows.

\end{proof}

\textbf{Proof of Theorem} \ref{thm:precomp-game-entropy-bound}. 
\begin{proof}

Since $u_{1}^{(\sigma_{pre},\Delta \tilde{\sigma}_2^{*})}=1$, $P_{norm}(v')=1$ for any $v' \leq 1$. Since the Nash equilibrium value is $v$, we must have $v \geq \tilde{u}^{(\Delta \tilde \sigma_1,\Delta \tilde{\sigma}_2^{*})}$ for any mixture of precomputation strategies $\Delta \tilde \sigma_1$ for player 1. Write $H(v'):=H\left(\mathcal{S}^{(\sigma_1,\Delta \tilde{\sigma}_2^{*}),\sigma_{pre}}(v')\right)$ as a shorthand, where we associate the mixed strategy $\Delta \tilde{\sigma}_2^{*}$ with a corresponding behavior strategy in the original game by Kuhn's theorem. Because the Nash equilibrium value of the game is $v$, any precomputation strategy for player $1$ must have utility $\leq v$ in the meta game. Theorem \ref{thm:rand-vs-precomp} provides the existence of a precomputation strategy for player $1$, and gives us that $\forall \epsilon \in (0,1), v' \in [0,1]$, 

\begin{align*}
    v &\geq (1-\epsilon)v' -(1-\epsilon)L \lambda_1 e^{H(v')/\epsilon}
\end{align*}

Requiring $v'>v+e^{-k}$, we can set $\epsilon=\frac{v'-v-e^{-k}}{v'-e^{-k}}$. Rearranging, simplifying, and assuming $e^{-k}<\lambda_1 L$ gives

\begin{align*}
    H(v') & \geq \frac{v'-v-e^{-k}}{v'-e^{-k}} \log \left( \frac{e^{-k}}{\lambda_1 L}\right)\\
    & \geq (v'-v-e^{-k}) \left(\log \left (\frac{1}{\lambda_1 L}\right)-k\right)\\
    &\geq (v'-v-0.01) \left(\log \left( \frac{1}{\lambda_1 L}\right)-5\right)\\
\end{align*}

for $k=5$. When $e^{-5}\geq \lambda_1 L$ the final inequality is non-positive and therefore automatically a lower bound on $H(v')$. 
\end{proof}

\begin{algorithm}[tb]
\caption{Compute best precomputation response.}\label{alg:greedy-precomp}
\textbf{Input}: $(\sigma_1,\sigma_2),\sigma_{pre},\lambda_1,\epsilon,\delta$\\
\textbf{Output}: A precomputation strategy $\tilde{\sigma}_1'$ for player 1 which\\ approximately maximizes $\tilde{u}^{(\tilde{\sigma}_1',\sigma_2)}$
\begin{algorithmic}[1]

\STATE Define $est(h,K)$ to be an estimate of $u_{1}^{\sigma_1,\sigma_2}(h)$ using the mean value of $K$ games sampled from history $h$ played according to $\sigma=(\sigma_1,\sigma_2)$. This can be computed in time $\mathcal{O}(LK)$.

\STATE Initialize $\text{BESTCHOICES},S = \emptyset$. 

\STATE Let $\text{bestvalue}(h)=est(h,K)\times \pi^{(\sigma_{pre},\sigma_2)}(h)$ if $\pi^{(\sigma_{pre},\sigma_2)}(h)<\lambda_1$. Otherwise, $\text{bestvalue}(h)=\max[\sum_{h' \in Succ(h)} \text{bestvalue}(h')-\lambda_1,$ $est(h,K)\pi^{(\sigma_{pre},\sigma_2)}(h)]$. If the left argument of the $\max$ is larger, we add $h$ to $\text{BESTCHOICES}$. Here we set $K=\mathcal{O}\left( \ln\left(\frac{A^2(L)}{\lambda_1}\right)/\epsilon^2\right)$. 

\STATE If the starting history $\emptyset \in \text{BESTCHOICES}$, then depth first search from $\emptyset$, where there is an edge $h \rightarrow h' \in Succ(h)$ iff $h' \in \text{BESTCHOICES}$. Add each history visited to $S$.

\STATE \textbf{Return} $\tilde{\sigma}_1'$, the precomputation strategy which memorizes $\sigma_{pre}$ on set $S$. 

\end{algorithmic}
\end{algorithm}

\textbf{Proof of Theorem} \ref{thm:find-precop-fixed}.

\begin{proof}
Let $S$ be the memorization set of the precomputation strategy $\tilde{\sigma}_1$. Then without loss of generality we can assume $h \in S \implies \pi^{(\sigma_{pre},\sigma_2)}(h) \geq \lambda_1$; otherwise, we could strictly increase $\tilde{u}^{(\tilde{\sigma}_1,\sigma_2)}_{1}$ by removing $h$ and any descendants of $h$ from $S$. The second observation is that we can conclude $|S| \leq \frac{L+1}{\lambda_1}$ from Lemma \ref{Lemma:bound-on-set-size}.

\

Algorithm 1 recursively computes an optimal precomputation memorization set $S$, but uses the approximation $est(h,K)$ in place of $u_1^{(\sigma_1,\sigma_2)}(h)$. Let $W_1=\{h \in H| \pi^{(\sigma_{pre},\sigma_2)}(h) \geq \lambda_1\}$, $W_2=Succ(W_1)$ and $W = W_1 \cup W_2$. If we have $\forall h \in W, |est(h,K)-u_1^{(\sigma_1,\sigma_2)}(h)| < \frac{\epsilon}{2}$, then if there exists some $\tilde{\sigma}_1$ with true value $\geq v$, the precomputation strategy $\tilde{\sigma}_1$ with precomputation set $S$ has value at least $v-\frac{\epsilon}{2}$ when we use the approximated utilities, and so Algorithm 1 finds some precomputation strategy with approximated value at least $v-\frac{\epsilon}{2}$. Conversely, this precomputation strategy has real value at least $v-\epsilon$. 

\ 

We have $|W| \leq |A|^2 \frac{(L+1)}{\lambda_1}$. Chernoff and union bound give us that the probability of the event $\exists h \in W$ s.t. $|est(h,K)-u_1^{(\sigma_1,\sigma_2)}(h)| \geq \frac{\epsilon}{2}$ is 

$$\leq |W| 2 \exp\left( -\frac{K}{2}(\epsilon/2)^2\right) \leq \delta$$

for $K=16 \ln \left(\frac{|W|}{\delta}\right)/\epsilon^2=16 \ln\left(\frac{A^2(L+1)}{\delta \lambda_1}\right)/\epsilon^2$. 

\ 

Thus with probability at least $1-\delta$, the estimates are all within $\frac{\epsilon}{2}$ of the true values and we find a precomputation policy $\tilde{\sigma}_1'$ with $\tilde{u}^{(\tilde{\sigma}_1',\sigma_2)}_{1} \geq v-\epsilon$. The algorithm considers at most $|W|$ histories, where each history takes time $L K$ to sample, giving a total running time of $|W| L K=\mathcal{O}\left(|A|^2 \frac{L^2}{\lambda_1\epsilon^2}  \ln(\frac{A^2L}{\lambda_1})\right)$. 
\end{proof}

\textbf{Proof of Lemma} \ref{Lemma:bound-on-set-size}.

\begin{proof}
We have 

$$L+1\geq \sum_{l=0}^L \sum_{\substack{h \in H \\ |h|=l}} \pi^{\sigma_1,\sigma_2}(h)=\sum_{h \in H} \pi^{\sigma_1,\sigma_2}(h)$$

$$\geq \sum_{\substack{h \in H \\ \pi^{\sigma_1,\sigma_2}(h) \geq p}} \pi^{\sigma_1,\sigma_2}(h) \geq p|\{h \in H | \pi^{\sigma_1,\sigma_2}(h) \geq p\}|$$

so 

$$|\{h \in H | \pi^{\sigma_1,\sigma_2}(h) \geq p\}| \leq \frac{L+1}{p}$$
\end{proof}

\begin{lem}\label{lem:bound_Wi}
Let $$W_{1}=\left\{h \in H|  \sup_{\Delta \tilde{\sigma}_{2} \in \Delta Pre(\sigma_{pre},\sigma_{2})} \pi^{\sigma_{pre},\Delta \tilde{\sigma}_{2}}(h) \geq \lambda_1\right\}$$

$$W_{2}=\left\{h \in H|  \sup_{\Delta \tilde{\sigma}_{1} \in \Delta Pre(\sigma_{pre},\sigma_{1})} \pi^{\Delta \tilde{\sigma}_{1},\sigma_{pre}}(h) \geq \lambda_2\right\}$$

$$Succ(W_i):=\cup_{h \in W_i} Succ(h)$$ for $i \in \{1,2\}$

$$W:=W_1\cup W_2 \cup W$$ 

Then $|W| \leq \mathcal{O}\left(A^2(L+2)^2(\frac{1}{\lambda_1}+\frac{1}{\lambda_2})\right)$, where $L=\max_{h \in H} |h|$ and $A=\max_{h \in H} |A(h)|$. 
\end{lem}

\textbf{Proof of Lemma} \ref{lem:bound_Wi}.

\begin{proof}
Let $\tilde{\sigma}_{-i,l}$ be the precomputation strategy for player $-i$ (the player opposing player $i$) which plays $\sigma_{pre}$ on every $h \in H$ such that $|h|<l$, and $\sigma_{-i}$ otherwise. Let $l(h)=\max_{|h|} \pi^{\sigma_{pre},\tilde{\sigma}_{-i,l}}(h) \geq p$, and $\infty$ if no such $l$ exists. Define 

$$W_i=\left\{h \in H|  \max_{\Delta \tilde{\sigma}_{-i} \in \Delta Pre(\sigma_{pre},\sigma_{-i})} \pi^{\sigma_{pre},\Delta \tilde{\sigma}_{-i}}(h) \geq p\right\}$$

Then we have

$$|W_i| = \sum_{l'=0}^{L+1} \sum_{\substack{h \in H \\l(h)=l'}} 1 \leq \sum_{l'=0}^{L+1} |h \in H:\pi^{\sigma_{pre},\tilde{\sigma}_{-i,l}}(h) \geq p|$$
$$\leq (L+2)\frac{(L+1)}{p} \leq \frac{(L+2)^2}{p}$$

The first equality follows because if $\pi^{\sigma_{pre},\Delta\tilde{\sigma}}_2(h) \geq p$ for some mixed precomputation strategy $\Delta\tilde{\sigma}_2$, then this must hold true for some pure precomputation strategy $\tilde{\sigma}_2$; if $\tilde{\sigma}_2$ precomputes on histories $h_{:0},\dots,h_{:l}$ for some $l$, then $\pi^{\sigma_{pre},\tilde{\sigma}_{-i,l}}(h) \geq p$. The last inequality follows from Lemma \ref{Lemma:bound-on-set-size}. Thus setting $p=\lambda_{-i}$, using that $|W_i \cup Succ(W_i)| \leq \mathcal{O}(A^2 |W_i|)$, and summing for $i=1,2$ gives the required bound. 
\end{proof}

\begin{lem}\label{lemma:extensive-form-game-equiv} (Kuhn's Theorem applied to the meta precomputation game). Given a two player game $G$ with policies $\sigma_1,\sigma_2,\sigma_{pre}$, define the extensive form imperfect information, perfect recall game $G'$ with the following tree structure:
\begin{enumerate}
    
    \item Each history $h' \in H'$ for game $G'$ corresponds to a history $h=Q(h') \in H$ in the original game $G$. At history $h'$, if $h'$ is not a chance node, then $P(h')=P(h)\in \{1,2\}$. Instead of player $P(h')$ choosing an action in $A(h)$, $P(h')$ chooses whether to continue precomputing if they have not already stopped (action$=1$) or stop precomputing (action=$0$). $h'$ then transitions to a chance node $h''$ ($P(h'')=c$), which picks an action $a \in A(h)$ according to the distribution $\sigma_{pre}(h)$ if $P(h')$ chose to precompute in the previous step, or $\sigma_{P(h)}(h)$ otherwise. The history which is transitioned to after the chance node in game $G'$ is then associated with history $\pi(h,a)$ in game $G$. 
    
    \item Players can see their own actions and the actions of the chance player, but not the actions of the opposing player. Formally, two histories $h',h'' \in H'$ for player $P(h)=P(h')=i\in \{1,2\}$ are in the same information set iff they have the same sequence of actions for player $i$ and the chance player leading up to them. 
    
    \item Terminal histories in game $G'$ are histories $h' \in H'$ where $P(h') \in \{1,2\}$ and the associated history $h \in H$ is a terminal history in game $G$. In each case, the utility of this terminal history $h'$ is $u'_1(h')=u_1^{(\sigma_1,\sigma_2)}(h)- \lambda_1 z_1(h')+\lambda_2 z_2(h')$ for some functions $z_i$ to be specified.

\end{enumerate}

Then for every pair of behavior strategies $(\tilde{\sigma}_1',\tilde{\sigma}_2')$ in $G'$ with utility $u$, there is a pair of mixed strategies $(\tilde{\sigma}_1=f(\tilde{\sigma}_1'),\tilde{\sigma}_2=f(\tilde{\sigma}_2'))$ in the meta precomputation game with the same utility $u$, and vice versa (for some function $f$).

\end{lem}

\textbf{Proof of Lemma} \ref{lemma:extensive-form-game-equiv}.

\begin{proof}
First, we argue that there is a bijective correspondence between pure strategies in the meta precomputation game $G$, and pure strategies in game $G'$. If we can establish this equivalence, then we know that there is also a bijective relationship between mixed strategies in $G$ and mixed strategies in $G'$. Finally, by Kuhn's Theorem, we know that there is a correspondence between mixed strategies in game $G'$, and behavior strategies in game $G'$ (because $G'$ is an extensive form perfect recall game), from which the conclusion of Theorem \ref{lemma:extensive-form-game-equiv} follows. It remains to show the precise relationship between pure strategies in the meta precomputation game $G$, and pure strategies in $G'$.

Fix any pure strategies $\tilde{\sigma}_{i}\in Pre(\sigma_{pre},\sigma_i)$ for $i \in \{1,2\}$ in the meta-precomputation game $G$, where $\tilde{\sigma}_{i}$ has memorization set $S_i$. Recall that for every $h' \in H'$ for the game $G'$ where $P(h') \not = c$, there is a unique associated history $Q(h'):=h$ in game $G$. We map $\tilde{\sigma}_{i}$ to the pure strategy $\tilde{\sigma}'_{i}$ in game $G'$, where player $i$ chooses action 1 (to precompute) at history $h'$ iff $Q(h') \in S_i$ and player 1 has not previously chosen to stop precomputing. Note that this is mapping is valid (i.e. respects information sets in game $G'$) and is bijective: any information set $I$ for player $i$ in game $G'$ where player $i$ has more than one choice (i.e. has always chosen action 1 leading up to this history) is uniquely determined by the sequence of actions of the chance player, which corresponds to a unique history $h$ in game $G$. Thus the inverse map, from any pure strategy $\tilde{\sigma}'_{i}$ in game $G'$ to a pure strategy  $\tilde{\sigma}_{i}$ in game $G$, where $h \in S_{i}$ if player $i$ chooses to precompute at the information set corresponding to history $h$ in game $G'$, is well defined. Finally, we need to show that

$$\tilde{u}_{1}^{(\tilde{\sigma}_{1},\tilde{\sigma}_{2})}=(u')_{1}^{(\tilde{\sigma}'_{1},\tilde{\sigma}'_{2})}$$

which establishes the desired relationship. We begin by defining the functions $z_i$ for $i \in \{1,2\}$. For $h' \in H'$, let  $\pi'_{c}(h')$ be the probability of reaching $h'$ where we only take into account the chance player (i.e. the product of the probabilities of all edges leading to $h'$ where the chance player chooses an action). Now fix any terminal history $h'\in Z'$ for game $G'$. Without loss of generality, we assume $\pi'_{c}(h') >0$ (otherwise we can remove it from the game without affecting the utility). Let $M_{i}(h')=\{h'_{i,1},\dots,h'_{i,j_{i}}\}$ be the set of histories $h'_{i,1} < h'_{i,2},\dots,h'_{i,j_{i}}<h'$ where player $i$ chose to memorize (action 1) in order to reach history $h'$. Then we let

$$z_{i}(h'):=\sum_{h'' \in M_{i}(h')} \frac{1}{\pi'_{c}(h'')}$$

Let $\tilde{\sigma}=(\tilde{\sigma}_{1},\tilde{\sigma}_{2})$ and $\tilde{\sigma}'=(\tilde{\sigma}'_{1},\tilde{\sigma}'_{2})$. If $Z'$ is the set of terminal histories for $G'$ where $\pi_{c}(h') \not = 0$, we can compute

\begin{align*}
    (u')_{1}^{\tilde{\sigma}'} & = \sum_{h' \in Z'} (\pi')^{\tilde{\sigma}'}(h')[u_{1}(Q(h'))\\
    & -\lambda_1 z_1(h')+\lambda_2 z_2(h')]\\
\end{align*}
Focusing on the first term gives
\begin{align*}
    \sum_{h' \in Z'} (\pi')^{\tilde{\sigma}'}(h')u_{1}(Q(h')) &= \sum_{h \in Z} \pi^{\tilde{\sigma}}(h)u_{1}(h)
\end{align*}

by construction of the game $G'$. Focusing on the remaining terms, we have 

\begin{align*}
    \sum_{h' \in Z'} (\pi')^{\tilde{\sigma}'}(h')z_i(h') &= \sum_{h' \in Z'} (\pi')^{\tilde{\sigma}'}(h')\sum_{h'' \in M_{i}(h')} \frac{1}{\pi'_{c}(h'')}\\
    &= \sum_{h' \in Z'} (\pi')^{\tilde{\sigma}'}(h')\sum_{\substack{h''<h'\\ Q(h'') \in S_i}} \frac{1}{\pi'_{c}(h'')}\\
    &=\sum_{h \in S_i} \sum_{\substack{h' \in Q^{-1}(h)\\ (\pi')^{\tilde{\sigma}'}(h')>0}} \pi'_{c}(h')\sum_{\substack{h'' \in Z' \\ h'' > h' \\ (\pi')^{\tilde{\sigma}'}(h'')>0}} \frac{1}{\pi'_{c}(h'')}\\
    &=\sum_{h \in S_i} \sum_{\substack{h' \in Q^{-1}(h)\\ (\pi')^{\tilde{\sigma}'}(h')>0}} \sum_{\substack{h'' \in Z' \\ h'' > h'\\ (\pi')^{\tilde{\sigma}'}(h'')>0} } \pi'_{c}(h',h'')\\
    &=\sum_{h \in S_i} \sum_{\substack{h' \in Q^{-1}(h)\\ (\pi')^{\tilde{\sigma}'}(h')>0}} 1\\
    &=\sum_{h \in S_i}1=|S_i|
\end{align*}

The first two equalities follow from the definitions of $z_i$ and $M_i$. In the third equality, we reverse the order of the summation. Instead of fixing a terminal history $h' \in Z'$ and summing over all ancestors $h''<h'$ where player $i$ chose to precompute, in the third equality we sum over all histories $h'$ where player $i$ chose to precompute (i.e. $h' \in Q^{-1}(h)$ where $h \in S_i$ and $h'$ has positive support on $\tilde{\sigma}'$), and then over the relevant terminal histories. Note that the summation $\sum_{\substack{h' \in Q^{-1}(h)\\ (\pi')^{\tilde{\sigma}'}(h')>0}}$ is over exactly one element (using the fact that we are only considering pure strategies). We are able to replace $(\pi')^{\tilde{\sigma}'}(h')$ by $\pi'_{c}(h')$ because $\tilde{\sigma}'$ consists of pure strategies, i.e. for any $h'$ in the support of $(\pi')^{\tilde{\sigma}'}$, the probabilities on the edges of the game tree leading up to $h'$ where players $1$ and $2$ make moves are all $1$, and so only the probability contribution from the chance player is relevant. 

Collecting these results, we find that
\begin{align*}
    (u')_{1}^{\tilde{\sigma}'} &=\sum_{h \in Z} \pi^{\tilde{\sigma}}(h)u_{1}(h)-\lambda_1 |S_1|+\lambda_2|S_2|\\
    &=\tilde{u}_{1}^{\tilde{\sigma}}\\
\end{align*}

showing the required equivalence.

\end{proof}

\begin{algorithm}[tb]
\caption{Compute an approximate Nash equilibrium to the meta-precomputation game.}\label{alg:find-equi}
\textbf{Input}: $(\sigma_1,\sigma_2),\sigma_{pre},\lambda_1,\lambda_2,\epsilon,\delta$\\
\textbf{Output}: An $\epsilon$-Nash equilibrium to the meta precomputation game. 
\begin{algorithmic}[1]

\STATE Define $est(h,K)$ to be an estimate of $u_{1}^{\sigma_1,\sigma_2}(h)$ using the mean value of $K$ games sampled from history $h$ played according to $\sigma=(\sigma_1,\sigma_2)$. This can be computed in time $\mathcal{O}(LK)$.

\STATE Define $G$ to be the meta-precomputation game.

\STATE Compute the set of high probability histories $W$ for $G$. Let $G'$ be the equivalent extensive form game for $G$ (Lemma \ref{lemma:extensive-form-game-equiv}), where histories $h'\in H'$ for game $G'$ are now made to be terminal histories if they are associated with low probability histories $h=Q(h') \in H$ for game $G$ (i.e. $h \not \in W$ from Lemma \ref{lem:bound_Wi}). If $h'$ is one of these terminal histories, we estimate the utility by $est(Q(h),K)+\lambda_1 z_1(h')-\lambda_2 z_2(h')$ (corresponding to Lemma \ref{lemma:extensive-form-game-equiv}) for $K=\mathcal{O}(\ln(\frac{|W|}{\delta})/\epsilon^2)$.

\STATE Run vanilla CFR from \cite{regretminCRF1} on this approximate game $G'$ for $\mathcal{O}\left(\frac{|W|^2}{\epsilon^2}\right)$ iterations.

\STATE Return the result of CFR on game $G'$.

\end{algorithmic}
\end{algorithm}

\textbf{Proof of Theorem} \ref{thm:CFR-on-game}

\begin{proof}
Suppose we pick $K=\mathcal{O}(\ln(\frac{|W|}{\delta})/\epsilon^2)$ samples for each call to $est(h,K)$, so that with probability at least $1-\delta$ we have $\forall h \in Succ(W_1)\cup Succ(W_2),|est(h,K)-u^{(\sigma_1,\sigma_2)}(h)|<\epsilon/x'$ for some constant $x'$ (by applying Chernoff+Union bound, similarly to the proof of Theorem \ref{thm:find-precop-fixed}). Let $\tilde{u}'_{1}$ be the utility of the extensive form game where we use the estimated values to compute the terminal values of the game. Suppose we run counter factual regret minimization to get a $2\frac{\epsilon}{x}$ Nash equilibrium $(\Delta \tilde{\sigma}_1,\Delta \tilde{\sigma}_2)$ to this game using utility function $\tilde{u}'_{1}$. Then we have

\begin{align*}
    \tilde{u}^{(\Delta \tilde{\sigma}_1,\Delta \tilde{\sigma}_2)}_{1}+\frac{\epsilon}{x'} &\geq (\tilde{u}')_{1}^{(\Delta \tilde{\sigma}_1,\Delta \tilde{\sigma}_2)}\\
    & \geq \sup_{\Delta \tilde{\sigma}_1 \in \Delta(Pre(\sigma_{pre},\sigma_1))} (\tilde{u}')_{1}^{(\Delta \tilde{\sigma}_1,\Delta \tilde{\sigma}_2)}-2\frac{\epsilon}{x}\\
    & \geq \sup_{\Delta \tilde{\sigma}_1 \in \Delta(Pre(\sigma_{pre},\sigma_1))} (\tilde{u})_{1}^{(\Delta \tilde{\sigma}_1,\Delta \tilde{\sigma}_2)}-2\frac{\epsilon}{x}-\frac{\epsilon}{x'}
\end{align*}

Thus 

$$\tilde{u}_{1}^{(\Delta \tilde{\sigma}_1,\Delta \tilde{\sigma}_2)}+2\left(\frac{\epsilon}{x'}+\frac{\epsilon}{x}\right) \geq \sup_{\Delta \tilde{\sigma}_1 \in \Delta(Pre(\sigma_{pre},\sigma_1))} (\tilde{u})_{1}^{(\Delta \tilde{\sigma}_1,\Delta \tilde{\sigma}_2)}$$

and the analogous inequality holds for player $2$, giving a $2\left(\frac{\epsilon}{x'}+\frac{\epsilon}{x}\right)$ Nash equilibrium. Picking $x'=x=4$ gives an $\epsilon$ Nash equilibrium. 

Each round of Counter Factual Regret Minimization requires time $\mathcal{O}(|W|)$, and after $T$ rounds the average regret for player $i$ is

$$R_{i} \leq \frac{|W|}{\sqrt{T}}$$

Setting $R_{i} \leq \frac{\epsilon}{x}$ to give a $2\frac{\epsilon}{x}$ Nash equilibrium gives

$$T \geq \mathcal{O}\left(\frac{|W|^2}{\epsilon^2}\right)$$

The total runtime for the regret minimization component is $\mathcal{O}(T \times |W|)=\mathcal{O}\left(\frac{|W|^3}{\epsilon^2}\right)$, and the runtime for sampling utility estimates is $\mathcal{O}(K |Succ(W_1)\cup Succ(W_2)| L)$, giving a total runtime bound of

$$\frac{1}{\epsilon^2}\left(A^2L^2\left(\frac{1}{\lambda_1}+\frac{1}{\lambda_2}\right)\right)^3$$

$$+\frac{1}{\epsilon^2}A^2L^3\left(\frac{1}{\lambda_1}+\frac{1}{\lambda_2}\right)\ln\left(AL\left(\frac{1}{\lambda_1}+\frac{1}{\lambda_2}\right)/\delta\right)$$

\end{proof}

\textbf{Experimental Details:}

Experiments were run on a single thread of a Intel Xeon E5-2678 v3 2.5GHz CPU for around 3 hours per plot. While it is feasible to run this experiment in full with more hardware, we make the following key simplifications to make the computation manageable on a desktop:

\begin{enumerate}
    \item Instead of considering all possible actions, we limit policies to play the from the top $k=2$ choices recommended by the Stockfish engine. 
    
    \item We set the maximum game length to $L=100$ half moves; a draw with utility $0.5$ is declared beyond this point.
    
    \item Instead of explicitly estimating the utility of each board state, we use Stockfish to compute the centerpawn (cp) score of each board state. If there is a centerpawn advantage of $\geq 400$ for any player, then this is calculated as a win for that player ($u=1$ or $0$). Otherwise the utility value is given as a draw ($u=0.5$). This choice (as opposed to e.g. estimating the expected value as a linear function of the cp score) was made to give the qualitative interpretation of the results a more conservative lower bound. In particular, we can interpret the utility of the precomputing player as a qualitative indication of the fraction of the time they are able to reach an overwhelming advantage against their opponent (where we can be reasonably sure they would win if playing normally from that point onwards). In contrast, it isn't clear that e.g. a 50cp score for white (scored by Stockfish(50ms)) would translate to Stockfish(10ms) having a slightly higher probability of winning, because Stockfish(10ms) may not be powerful enough to make use of that advantage. 
\end{enumerate}